\documentstyle[11pt,newpasp,twoside,epsf]{article}
\def\edcomment#1{\iffalse\marginpar{\raggedright\sl#1\/}\else\relax\fi}
\reversemarginpar

\begin{document}
\title{The Effect of Heating and Cooling on Scaling Laws of X-Ray Clusters}
 \author{Orrarujee Muanwong}
\affil{Astronomy Centre, University of Sussex, Brighton, BN1 9QJ, UK}

\begin{abstract}
We present results of including heating and cooling in cosmological
simulations of the $\Lambda$CDM cosmology and demonstrate their effects
on scaling laws of galaxy clusters. The scaling relations when radiative
cooling is included are in good agreement with observations but the fraction
of cooled gas is on the upper limit allowed by observations. On the contrary,
the preheating model has a more realistic cooled fraction but the scaling
relations are less well reproduced. 

\vskip -1.0truecm
\end{abstract}

\section{Introduction}
The predicted X-ray luminosity and temperature
relation from self-similarity (${L_X \propto T^{2}_X})$ is shallower than
the observed one (${{L_X}\propto T^{\sim3}_{X}}$). 
The steepening of the relation implies that low temperature 
clusters are less luminous 
than they would have been if self-similarity were to hold. A physical
explanation of this effect is that as the luminosity is roughly 
proportional to
the gas density squared, there must exist processes which lower the density
preferentially in low temperature clusters, hence the steepening of the
relation. Such processes are for example preheating and radiative cooling
of the gas. 
 
Ponman, Cannon, \& Navarro (1999) have shown that the observed core entropies of low
temperature clusters are in excess of what can be achieved by gravitational
collapse alone. 
They suggest that preheating can raise the entropy before the infall and 
that level of entropy is maintained during gravitational collapse. 
As the entropy scales as ${T/n^{2/3}_e}$, where
${n_{e}}$ is the electron gas density, and the temperature of the
gas is fixed by the virial temperature of the halo, 
the excess entropy implies a
decrease in density, hence a decrease in luminosity. Energy sources 
which can inject energy into the intergalactic medium (IGM) are 
for example AGN, quasars, hypernova explosions,
supernova explosions and so on. However, it is still debatable whether the
energy injection efficiency of the heating sources is sufficient to
provide the required entropy level.

On the other hand, radiative cooling is a process which is also important
throughout the history of cluster formation. Low entropy gas, i.e. dense
and cold gas, is removed from the ICM, leaving behind high entropy material.
Pearce et al (2000) have shown that when radiative cooling is included, the
gas is slightly heated and the luminosity is greatly reduced thus acting
so as to reconcile simulations and observations. 

In these Proceedings, we present results from cosmological simulations
of the $\Lambda$CDM cosmology in which 3 different models are considered.
In the first, 'non-radiative' model, non-adiabatic 
heating comes about only via shocks 
following gravitational collapse and the gas is not allowed to cool. 
The second 'cooling' model introduces radiative cooling such that 20
percent of the gas in the box has turned into stars by the current
day - a value that is higher than suggested by observations (Balogh et
al 2001). Finally, the third 'preheating' model has both radiative cooling 
and an impulsive heating event in which 0.1 keV per particle is injected
into the IGM at a redshift of 4. This lowers the cooled fraction down
to 10 percent.

The  simulations consist of 160$^{3}$  particles  each of gas and dark
matter  within  boxes  of side 100    $h^{-1}$  Mpc. The  cosmological
parameters are : density  parameter, $\Omega_{0}$ = 0.35; cosmological
constant,   $\Lambda_{0}$=0.65,   Hubble    parameter, $h=H_{0}$/100km
s$^{-1}$  Mpc$^{-1}$    =   0.71   and  baryon      density parameter,
$\Omega_{b}h^{2}$ = 0.019.  The metallicities  are varied with time in
the form of  $Z=0.3(t/t_{0}) Z_{\sun}$, where  $t/t_{0}$ is the age of
the Universe in units of the current time. The objective of this is to
mimic the gradual enrichment of the ICM by stars. Cluster
catalogues are constructed at redshift   zero. The catalogues are
complete down to   a  cluster mass  of  $1.18  \times 10^{13}  h^{-1}$
M$_{\sun}$ which corresponds to 500 particles of each species. Details
of cluster selection criteria can be found in Muanwong et al (2001).

\section{Results}

In  this section, cluster properties are compared to  those
observed.  We  calculate    the X-ray  properties   (temperature   and
luminosity) in a  soft X-ray band (0.3-1.5  keV) by using  the cooling
tables of Raymond \& Smith (1977). The luminosity is then corrected to
a bolometric  one by assuming  that  the gas is  isothermal.  The main
effect of using band-limited rather than bolometric fluxes is to raise
the emission-weighted temperature of low-mass clusters slightly.

\subsection{${{T_{X}}-M_{200}}$ Relations}

We compare the derived  ${{T_{X}}-M_{200}}$ relations of the simulated
catalogues  with  a    compilation of  observations made   by  Horner,
Mushotzky,  \& Scharf  (1999).  By  using  mass estimates from  galaxy
velocity  dispersions,  X-ray  temperature   profiles,  the isothermal
$\beta$-model  and  the surface   brightness deprojection method, they
obtain different scaling laws shown as dashed lines in Figure 1.

The  normalisation  of the cooling   simulation (triangles) is greater
than that of the non-radiative simulation  (squares) by about a factor
of 2 at the mass of 10$^{14} h^{-1}$  M$_{\sun}$. For clarity, we omit
the preheating  results but they lie roughly  between the two. Most of
the radiative cooling and preheating  clusters, particularly the small
systems, are hotter than  their  virial temperatures (solid line),  in
accord with expectation.   Non-radiative cooling  clusters  are cooler
compared to the virial  temperatures due to emission-weighting of cold
and dense  gas remaining in   the cluster core. The   slopes of all  3
simulated catalogues   are roughly  0.54.  We  find  a good  agreement
between  the    relations  from the   radiative   cooling  model   and
observations, given the spread of the 4 estimates from observations.

\begin{figure}[ht]  
\centering
\vskip -1.5truecm \plotone{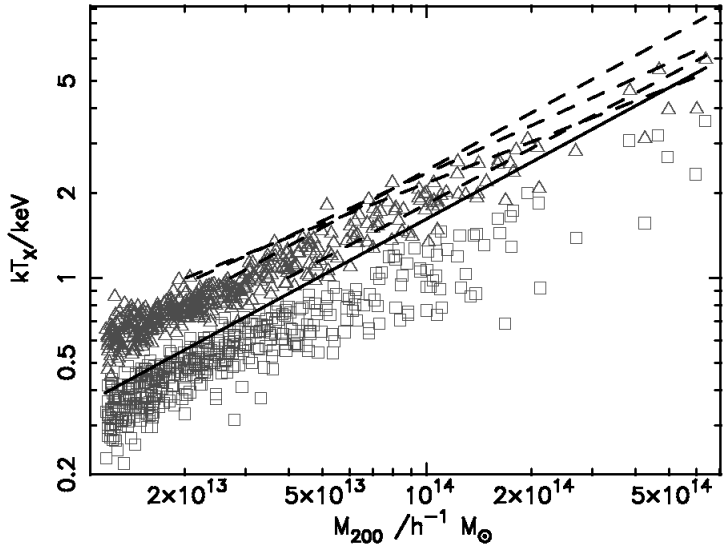}
\vskip -1.5truecm
\caption{The  temperature-mass  relation. The   dashed lines  are from
observations (Horner et   al  1999). The solid  line  is  the relation
between the virial temperature  and mass.  Non-radiative and radiative
cooling clusters are shown as squares and triangles, respectively.}
\vskip -0.5truecm
\end{figure}

\subsection{${L_{bol}-T_{X}}$ Relations}

The effect of radiative  cooling is demonstrated dramatically in terms
of luminosity.  The  ${L_{bol}-T_{X}}$ relations of the  non-radiative
cooling, radiative cooling  and preheating  simulations are shown   as
squares, triangles and circles  respectively  in Figure 2.  The  large
decrease in density  as a result of radiative  energy loss of  the gas
manifests  in a significant reduction  in luminosity. At a temperature
of 1 keV, the luminosity is reduced by a  factor of 100. The radiative
cooling model    again   is found to   have   a  good  agreement  with
observations as  shown as solid lines (Xue  \& Wu 200).  The different
solid   lines are  different fits   to their   subsamples, i.e. groups
(roughly below 1 keV), clusters (above 1 keV) and a mixture of the two
subsamples.

The  preheating model produces a  relation which lies  between that of
the  other two models.  Preheating   still results  in an increase  in
entropy,  causing the normalisation of the  relation to be a factor of
10 smaller than that of  the non-radiative cooling simulation, but the
agreement  with observations is not as  good as  the radiative cooling
simulation.   The     non-radiative clusters  are    consistent   with
self-similarity.  The   relations  in   the  other   two    models are
significantly  steeper   than  2, even  more   so  in low  temperature
clusters.

\begin{figure}  
\vskip -1.5truecm \centering \plotone{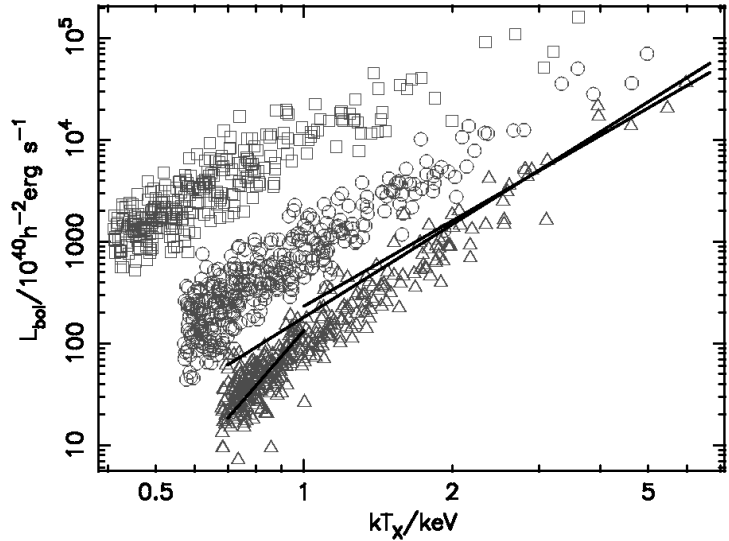}
\vskip -1.5truecm
\caption{The luminosity-temperature relation. The solid lines are from
observations (Xue  \& Wu 2000). Symbols are  as same  as those used in
Figure 1. Circles represent preheated clusters.}
\vskip -0.5truecm
\end{figure}

\section{Conclusions}
We have  demonstrated the   effect of heating   and cooling  on  X-ray
properties of clusters in  cosmological simulations. The
radiative  cooling model can   reproduce both the temperature-mass and
luminosity-temperature       relations  in    good  agreement     with
observations. However,  the   fraction of cooled  gas  is considerably
higher  than observed. In contrast, the  preheating model results in a
lower cooled  fraction but gives a  poor match to the observed scaling
relations.  As  a possible resolution, we are currently undertaking a
simulation which includes energy feedback from supernovae.

\section{Acknowledgement}
I am grateful for the financial support from the Thai Government. I wish to
thank Peter Thomas and Scott Kay for reading this manuscript and their
feedbacks.
Simulations presented here were carried out at Edinburgh Parallel 
Computing Centre as part of the Virgo Supercomputing Consortium.

\end{document}